\documentclass[aps,prl,groupedaddress,amssymb,amsmath,twocolumn,showpacs,superscriptaddress]{revtex4-1}

\begin{document}


\title{Logarithmic superdiffusivity of the $2$-dimensional anisotropic KPZ equation}

\author{Giuseppe Cannizzaro}
\affiliation{University of Warwick, 
Department of Statistics, 
CV4 7AL, Coventry
UK}

\author{Dirk Erhard}
\affiliation{Universidade Federal da Bahia, Instituto de Matem\'atica, Av. Adhemar de Barros, s/n, Ondina, Salvador - BA, 40170-110, Brazil}

\author{Fabio Lucio Toninelli}
\affiliation{Technische Universit\"at Wien,
Institut f\"ur Stochastik und Wirtschaftsmathematik, 
Wiedner Hauptstra\ss e 8-10, 
A-1040 Vienna,
Austria
}




\date{\today}

\begin{abstract}
  We study an anisotropic variant of the two-dimensional
  Kardar-Parisi-Zhang equation, that is relevant to describe 
  growth of vicinal surfaces and has Gaussian,
  logarithmically rough, stationary states.  While the folklore belief
  (based on one-loop Renormalization Group) is that the equation has
  the same scaling behaviour as the (linear) Edwards-Wilkinson
  equation, we prove that, on the contrary, the non-linearity induces the
  emergence of a logarithmic super-diffusivity. This phenomenon 
  is similar in flavour to the super-diffusivity for two-dimensional fluids
  and driven particle systems.
\end{abstract}



\maketitle


Stochastic growth phenomena are ubiquitous in non-equilibrium
statistical physics \cite{barabasi1995fractal}.  Over the last 20 years most of the attention has
focused on one-dimensional ($1d$) growing interfaces (e.g.  the
boundary of a bacterial colony spreading in a two-dimensional
medium). Experimental, theoretical and mathematical results succeeded in
unveiling the universal features (most notably, scaling exponents and
non-Gaussian limiting distributions) of what is by now known as the
$1d$ KPZ universality class. Also in dimension $d\ge3$ 
progress was made in both the physics and mathematics literature and 
recently the prediction \cite{Kardar} of
asymptotically Gaussian behaviour for small coupling constant has been
rigorously established. Instead, the harder case of $2d$ growth, on
which we focus here, is still to a large extent unexplored. We study
an anisotropic version of the $2d$ KPZ equation for which we determine
super-diffusive behaviour, contradicting the claim of diffusivity repeatedly made
in the previous literature.

The KPZ equation is the stochastic partial differential equation  
\begin{equation}\label{eq:KPZ}
\partial_t H= \frac12 \Delta H + \lambda \langle \nabla H, Q\nabla H\rangle +  \xi\,,
\end{equation}
where $H=H(t,x)$  depends on time $t\geq 0$ and
on a $d$-dimensional space coordinate $x$, $\Delta$ is the $d-$dimensional Laplacian, $\xi$ is the space-time
Gaussian white noise, i.e. $ \mathbb E{\xi(x,t)}=0$,
$\mathbb E({\xi(x,t)\xi(y,s)})=\delta (x-y)\delta(t-s)$  ($\mathbb E(\cdots)$ denoting the average), 
$Q$ is a  fixed $d\times d$ symmetric matrix and $\lambda\ge0$ tunes the strength of
the non-linearity (here, $\langle\cdot,\cdot\rangle$ denotes the usual scalar product in $\mathbb R^d$). 

The equation was introduced in a seminal
paper by Kardar, Parisi and Zhang~\cite{Kardar}, that focused on the situation in which $Q$
is the identity matrix, thus reducing the non-linearity to $|\nabla
H|^2$. In this case \eqref{eq:KPZ} is also connected to the partition function $Z$ of a
$(d+1)$-dimensional directed polymer in a random potential (the time
variable is  the $(d+1)$-th space coordinate) via the transformation $Z=\exp (\lambda H)$. 
More generally,~\eqref{eq:KPZ} serves as a model for $(d+1)$-dimensional
stochastic growth, the non-linear term encoding the slope-dependence of the growth mechanism, 
and it is presumed to arise as the scaling limit of a large class of interacting particle systems. 
The phenomenological connection with microscopic growth models is the
following: for \eqref{eq:KPZ} to correctly describe the height
fluctuation process around a macroscopically flat state of slope
$\rho\in \mathbb R^d$, one should take $Q=D^2v(\rho)$, where $v(\rho)$
is the average speed of growth and $D^2 v$ is the Hessian matrix of $v$. 

A natural problem associated to~\eqref{eq:KPZ} is to determine whether the
non-linearity is relevant or not in a Renormalization Group (RG) sense, i.e. whether large-scale
features of the equation, such as roughness and growth exponents $\alpha,\beta$,
differ or coincide with those of the linear Edwards-Wilkinson (EW) equation corresponding to~\eqref{eq:KPZ} with
$\lambda=0$. It has been argued in~\cite{Kardar} and confirmed since
then in many  works \cite{SS, ACQ, BQS11,  Virag, QS1} that the non-linearity is  relevant in dimension $d=1$ (the growth exponent changes from $\beta_{EW}=1/4$ to $\beta_{KPZ,d=1}=1/3$),
whereas it is not if $d\geq 3$, provided $\lambda$ is smaller than a
critical threshold
$\lambda_c(d)$ (the mathematical proofs of this \cite{Magnen,Gu2018b,CCM2} require that $Q=\mathbb I$). 
In $2$ dimensions, however, the situation is more subtle: the
non-linearity is dimensionally marginal and the qualitative behaviour
of~\eqref{eq:KPZ} was predicted in~\cite{W91,barabasi1995fractal} to
depend on the sign of the determinant of $Q$. In the case of
$\det Q>0$ the non-linearity changes the growth and roughness exponent (see e.g. \cite{halpin20122+})
to two universal values
$\alpha_{KPZ,d=2}\approx 0.39...,\beta_{KPZ,d=2}\approx 0.24...$,
compatible with the exact scaling relation $\alpha+z=2$, with
$z=\alpha/\beta$ the dynamic critical exponent.  Instead, for
$\det Q\leq 0$, which is called ``anisotropic KPZ'' (AKPZ) and
includes both the linear equation $Q=0$ as well as models of growth of
vicinal surfaces \cite{W91}, the exponents should be the same as for
the EW equation, i.e.  $\alpha_{EW,d=2}=\beta_{EW,d=2}=0$, with
logarithmic instead of power-like fluctuation growth.  This has been
conjectured on the basis of one-loop RG
computations~\cite{W91,barabasi1995fractal} and supported by numerical
simulations \cite{Healy} of a discretized version of \eqref{eq:KPZ}.
Further, it has been claimed \cite{W91,barabasi1995fractal,Healy} that
the large-scale fixed point of \eqref{eq:KPZ} is the EW equation.  The
purpose of the present work is to disprove the latter claim. Indeed
our main result is that if $d=2$ and $Q={\rm diag}(+1,-1)$ is the
diagonal matrix with entries $(+1,-1)$ then, as soon as $\lambda\ne0$,
\eqref{eq:KPZ} is \emph{logarithmcally super-diffusive}, namely the
correlation length $\ell(t)$ behaves like
$\sqrt t\times (\log t)^{\tfrac{\delta}{2}}$ as time grows, for some
$\delta>0$, while EW has the usual diffusive growth
$\ell(t)\sim \sqrt t$. Interestingly, the exponent $\delta$ does not
continuously go to zero as $\lambda\to0$ and, in fact, a mode-coupling
theory computation suggests that $\delta=1/2$ for every $\lambda\ne0$.
A more precise statement of the results, together with an idea of the
proof, is given below. A full mathematical proof can be found in
\cite{CET}.  Before we proceed, let us remark that, even though in the
context of $2d$ growth our findings were unexpected, logarithmic
corrections to the diffusive scaling have already been observed (and
rigorously proved) for other two-dimensional out of equilibrium
systems such as driven particle systems (see \cite{BKS85, Yau} for the
asymmetric simple exclusion process, in which case though the value of
$\delta$ is $2/3$) and fluid models (see~\cite{Adler, LRY}) where
$\delta=1/2$.
%
%
%
%
%
%
%
%
%

A distinguishing feature~\cite{da2003nonlinear} of the $2d$ equation \eqref{eq:KPZ} with $Q={\rm diag}(+1,-1)$, i.e. the AKPZ equation
\begin{equation}\label{eq:KPZq}
\partial_t H= \frac12 \Delta H + \lambda [(\partial_{x_1} H)^2-(\partial_{x_2} H)^2] +  \xi
\end{equation}
is that it has a Gaussian log-correlated stationary state $\eta$. More
precisely, $\eta$ is a zero-mean Gaussian field (GFF in the
mathematical jargon) whose covariance is
$\mathbb E({ \eta(x)\eta(y)})\sim \log|x-y|$ (with $x=(x_1,x_2)$),
showing a vanishing roughness exponent. Note that the
stationary state is independent of $\lambda$.  As remarked
in~\cite{da2003nonlinear}, \eqref{eq:KPZq} is the only version of the
$2d$ KPZ equation (up to rotations) whose stationary state is
Gaussian.

The equation \eqref{eq:KPZq} is mathematically ill-posed: the solution
at fixed time is a GFF, that is merely a distribution, so that the
square $(\partial_{x_i} H)^2$ does not make sense. A usual way out
(that was already adopted implicitly in \cite{Kardar}) is to
regularize the equation. In \cite{CET}, we replaced
$(\partial_{x_i} H)^2$ by $\Pi((\Pi \partial_{x_i} H)^2$, where $\Pi$
is a cut-off in Fourier space, that removes all modes $|k|\ge 1$. The
non-linearity then becomes
$\mathcal{N}(H)= \Pi((\Pi \partial_{x_1} H)^2- (\Pi \partial_{x_2}
H)^2)$. As  observed
in \cite{CES19}, the stationary state of the regularized equation is
still the GFF $\eta$ and from now on we work with the stationary
process with initial condition $H(0)=\eta$. We expect that our results would hold unchanged if we
regularized the noise instead, as is often done.  Also,  in \cite{CET} we
work on a torus of side length $2\pi N$ instead of the infinite plane,
and $N$ is sent to infinity before any other limit is taken. For
lightness, we drop the $N$-dependence in the formulas below. 

A convenient way of encoding the growth in time of the correlation
length is through the \emph{bulk diffusion coefficient} $D_{bulk}(t)$
\cite{spohn2012large}.  For the KPZ equation, the usual way to define it is as in \cite{BQS11}. In our context, we let
$U=(-\Delta)^{1/2} H$ (this operation just means that, in Fourier
space, each Fourier mode $\hat U(t,k)$ is given by $|k|\hat H(t,k)$),
that solves a $2d$ stochastic Burgers equation whose
stationary state is simply the Gaussian white noise $\rho$, with
$\mathbb E{\rho(x)}=0,\mathbb E({\rho(x)\rho(y)})=\delta(x-y)$.  Then, $D_{bulk} $ reads
\begin{equation}
D_{bulk}(t)= \frac{1}{2t}\int_{\mathbb R^2} |x|^2S(t,x)\, dx\,,
\end{equation}
with
\begin{equation}
S(t,x)= \mathbb E({U(t,x) U(0,0)})\,.
\end{equation}
Note that $S(0,x)=\delta(x)$ and $t\times D_{bulk}(t)$ measures the spread
of correlations in time in a mean-square sense. The explicit solution of the EW equation yields that 
 $D^{EW}_{bulk}(t)=1$ independently of $t$, corresponding to the usual $\sqrt t$ growth of correlation length.
 Our main result is that in contrast, as soon as $\lambda\ne0$, there exists  $0<\delta\le 1/2$ such that 
\begin{equation}\label{maineq}
(\log t)^\delta \leq D_{bulk}(t) \leq (\log t)^{1-\delta}\,
\end{equation}
for $t$ large (to be precise, \eqref{maineq} is proven in the sense of
Laplace transforms, see \eqref{maineqL} below).  While we do not pin down
the precise value of $\delta$, we can prove that \emph{$\delta$ does
  not tend to zero as $\lambda\to0$}, while as mentioned it equals
zero for $\lambda=0$.  The result can be reformulated by saying that
the dynamic exponent $z$ is still $z=2$ like for EW, but the effect of
non-linearity changes the power-law behaviour by a non-trivial
logarithmic factor.

Another natural question for stochastic growth processes is how they
behave under rescaling. The $2d$ EW stationary equation is
well known to be scale-invariant under diffusive scaling, i.e.
\begin{equation}
H^\varepsilon (t,x):= H(t/\varepsilon^2, x/\varepsilon)
\end{equation}
has the same law as $H(t,x)$. Our second result shows that, for the
non-linear equation \eqref{eq:KPZq} with $\lambda\ne0$, this is not
true, not even asymptotically for $\varepsilon\to0$.  Namely, we
prove that the fields $H^\varepsilon(t,\cdot)$ and
$H^\varepsilon(0,\cdot)$ already decorrelate at times of order
$|\log \varepsilon|^{-\delta}\ll 1$ for $0<\delta\le 1/2$ as
above. We quantify this by verifying (\cite[Th. 1.2]{CET}) that given a smooth test
function $\varphi$ and letting $H^\varepsilon_\varphi$ be the centered
random variable
$H^\varepsilon_\varphi(t)=\int_{\mathbb R^2} {\rm d} x
\varphi(x)H^\varepsilon(t,x)$, the normalized covariance
\begin{eqnarray}
  \frac{{\rm Cov}({H^\varepsilon_\varphi(t),H^\varepsilon_\varphi(0)})}{{\rm Var}(H^\varepsilon_\varphi(0))}
\end{eqnarray}
is strictly smaller than $1$ for
$t\approx |\log \varepsilon|^{-\delta}\ll1$, \emph{uniformly as
  $\varepsilon\to0$}.  This result again indicates that the large
scale behaviour of~\eqref{eq:KPZq} differs from that of EW.

We emphasize that the above \emph{does not contradict} the numerical
findings of \cite{Healy}, but only its conclusion that the solution of
\eqref{eq:KPZq} shows a ``very rapid, unrelenting and nearly immediate
crossover to the EW fixed point''. In fact, \cite{Healy} numerically
observes $\sqrt{\log t}$ growth of fluctuations in time for \eqref{eq:KPZq}, which is the same growth as for EW:  this finding is in
agreement with rigorous results, see Theorem 1.5 in \cite{CET}, but it
does not address the question of logarithmic corrections to the diffusive
scaling or to $D_{bulk}$, which turns out to be the feature that
really distinguishes between the EW and AKPZ equations.

Before explaining how we prove \eqref{maineq}, let us briefly give a
heuristics, based on a mode-coupling approximation \cite{BKS85,S14}, which moreover leads to the conjecture $\delta=1/2$.
Let $\hat S(t,k)=(2\pi)^{-2} \mathbb E(\hat U(t,k) \hat U(0,-k)), k=(k_1,k_2)$ 
be the Fourier transform of $S$. A direct computation shows that $\hat S$ satisfies
the exact identity (see \cite[App. B]{CET} for details)
\begin{widetext}
\begin{align}
  (\partial_t+\frac{|k|^2}2)\hat S(t,k)=& - \frac{|k|^2\lambda^2 }{(2\pi)^6}\int_0^t {\rm d}s\,  e^{-\tfrac{|k|^2}{2}(t-s)}\int {\rm d}p \int {\rm d}q K_{p,k-p}K_{q,-k-q}\mathbb E\left[\hat U(s,p) \hat U(s,k-p)\hat U(0,q)\hat U(0,-k-q)
  \right]\label{e:Prod}
\end{align}
\end{widetext}
where $K_{p,q}=(p_2 q_2-p_1 q_1)/(|p| \, |q|)$ 
comes from the Fourier representation of the non-linearity in \eqref{eq:KPZq} 
and the integration is over the two-dimensional momenta $p,q$ subject to the conditions $|p|,|q|,|k-p|,|k+q|\le1$ 
due to the Fourier regularisation induced by $\Pi$. To get an (approximate) closed equation for $\hat S$, 
we perform a Gaussian approximation which allows to replace  
the average in~\eqref{e:Prod} by a Gaussian one. 
The conservation of momentum then readily implies that the contractions 
$\mathbb E[\hat U(s,p) \hat U(s,k-p)],\, \mathbb E[\hat U(0,q)\hat U(0,-k-q)]$ multiplied by $|k|^2$
do not contribute, and we obtain 
\begin{align}\label{e:ClosedEq}
  (\partial_t+\frac{|k|^2}2)\hat S&(t,k)=-\frac{2|k|^2\lambda^2}{(2\pi)^4} \int_0^t {\rm d} s\, e^{-\tfrac{|k|^2}{2}(t-s)}\nonumber\\
  &\times\int {\rm d}p (K_{p,k-p})^2\hat S(s,p)\hat S(s,k-p)\,. 
\end{align}
We now make the Ansatz
\begin{equation}\label{e:Ansatz}
\hat S(t,k) = \hat S(0,0)e^{-\tfrac{|k|^2}{2}t-c|k|^2t(\log t)^\delta}\,,
\end{equation}
for $k$ small and $t$ large. Notice that in this regime $k-p\approx p$, which means $(K_{k-p,p})^2\approx 1$, 
and $\exp(-|k|^2(t-s)/2)\approx 1$.  
Hence, computing the left and right hand side of~\eqref{e:ClosedEq} with $\hat S$ as in~\eqref{e:Ansatz} 
and then equating them, we are led to 
\begin{equation*}
-|k|^2 (\log t)^\delta \approx -|k|^2\lambda^2 (\log t)^{1-\delta}
\end{equation*}
which imposes the choice $\delta=1/2$. 
\medskip

%
%
%


%
%
%
%
%

The actual proof of~\eqref{maineq} given in~\cite{CET} starts by rewriting the bulk diffusion coefficient 
in its Green-Kubo formulation 
\begin{equation}
\begin{aligned}
D_{bulk}(t)=1+\frac{2\lambda^2}{t}\mathbb{E}\Big[\Big(\int_0^t{\rm d} s\, \overline{\mathcal{N}}(U(s))\Big)^2\Big] 
\end{aligned}
\end{equation}
where $\overline{\mathcal{N}}(U(s))$ is the spatial average of 
$\mathcal{N}(H(s,\cdot))=\mathcal{N}((-\Delta)^{-1/2}U(s,\cdot))$. 
Now, thanks to \cite[Lemma 5.1]{CES19} and the fact that $U$ is a stationary Markov process 
whose law at every fixed time is that of the spatial white noise $\rho$, 
the Laplace transform in $t$ of $t\times D_{bulk}(t)$, which we denote by $\mathcal{D}_{bulk}$, can be written as 
\begin{equation}\label{e:LDbulk}
\mathcal{D}_{bulk}(\mu)=\frac{1}{\mu^2}+\frac{1}{\mu^2}\mathbb{E}[\overline{\mathcal{N}}(\rho) (\mu-\mathcal{L})^{-1}\overline{\mathcal{N}}(\rho)]
\end{equation}
with $\mathcal{L}$ the generator of the Markov process $U$ and where
the expectation is taken with respect to the law of the stationary state $\rho$. 
In the Laplace transform sense,~\eqref{maineq} for large $t$ is equivalent to 
\begin{equation}\label{maineqL}
\frac{1}{\mu^2}|\log \mu|^\delta\leq\mathcal{D}_{bulk}(\mu)\leq \frac{1}{\mu^2}|\log \mu|^{1-\delta}
\end{equation}
for $\mu$ small. 

To have a better understanding of the expectation in \eqref{e:LDbulk}, recall that 
the bosonic Fock space associated to $\rho$  can be decomposed as $\oplus_{n\ge0} \Gamma L^2_n$, 
the $n$-particle sector $\Gamma L^2_n=L^2_{sym}(\mathbb{R}^{2n})$
being the space of square integrable functions which are symmetric in their $n$ 
two-dimensional coordinates, endowed with the usual $L^2$-scalar product $\langle\cdot,\cdot\rangle_n$. 
Let us remark that, denoting by $\mathfrak{n}$ the representation in Fock space of $\overline{\mathcal{N}}(\rho)$, that belongs to 
$\Gamma L^2_2$ since $\overline{\mathcal{N}}(\rho)$ is quadratic in $\rho$. 
Let $P_n$ be the orthogonal projection onto $\Gamma L^2_{\leq n}=\oplus_{j\leq n} \Gamma L^2_j$ and set 
$\mathcal{L}_n=P_n \mathcal{L} P_n$. 
It turns out (see~\cite[Lemma 3.1]{CET}) that 
the sequence $b_j(\mu)=\langle\mathfrak{n},(\mu-\mathcal{L}_j)^{-1}\mathfrak{n}\rangle_2$, satisfies 
\begin{equation}
b_3(\mu)\leq b_5(\mu)\leq\dots\leq b_4(\mu)\leq b_2(\mu)
\end{equation}
and 
\begin{equation*}
\lim_{j\to\infty} b_j(\mu)=b(\mu):=\langle\mathfrak{n},(\mu-\mathcal{L})^{-1}\mathfrak{n}\rangle_2\,,
\end{equation*}
where the right hand side equals  the expectation in~\eqref{e:LDbulk}. 
Therefore, in order to prove~\eqref{maineq}, it suffices to determine suitable 
upper and lower bounds for $b_{2j}(\mu)$ and $b_{2j+1}(\mu)$, respectively. 
To do so, note first that the symmetric part of $\mathcal{L}$, $\mathcal{L}_0$, acts in Fock space as 
$-\tfrac12\Delta$ so that in particular it leaves $\Gamma L^2_n$
invariant, i.e. it conserves the particle number. 
On the other hand, the antisymmetric part $\mathcal{A}$ can be written as the sum of 
$\mathcal{A}_+$ and $\mathcal{A}_-$, which are such that $-\mathcal{A}_+^*=\mathcal{A}_-$ and 
the former maps $\Gamma L^2_n$ to $\Gamma L^2_{n+1}$ while the latter to $\Gamma L^2_{n-1}$. 
If we recursively define the operators $\mathcal{H}_j$'s as
\begin{equation}\label{e:H}
\begin{aligned}
\mathcal{H}_3&=-\mathcal{A}_-(\mu-\mathcal{L}_0)^{-1}\mathcal{A}_+\\
\mathcal{H}_j&=-\mathcal{A}_-(\mu-\mathcal{L}_0+\mathcal{H}_{j-1})^{-1}\mathcal{A}_+
\end{aligned}
\end{equation}
we obtain the alternative representation 
\begin{equation}\label{e:b}
b_j(\mu)=\langle \mathfrak{n},(\mu-\mathcal{L}_0+\mathcal{H}_j )^{-1}\mathfrak{n}\rangle_2\,.
\end{equation} 
From~\eqref{e:H} it is immediate to verify that, for all $j$, $\mathcal{H}_j$ leaves each of the $\Gamma L^2_n$'s
invariant. In order to treat the inverse of  $\mu-\mathcal{L}_0+\mathcal{H}_{j-1}$ 
and get meaningful estimates for $b_j$, we need to control the  
$\mathcal{H}_j$'s in terms of explicit multiplication operators which act diagonally in momentum space, 
as $\mathcal{L}_0$ does. 
Thanks to the structure of~\eqref{e:H}, we can iteratively bound the $\mathcal{H}_j$'s 
starting from $\mathcal{H}_3$ and ultimately attain 
\begin{align}
\mathcal{H}_{2j+1} &\lesssim C^{2j+1}(-\mathcal{L}_0) \frac{\log(1+(\mu-\mathcal{L}_0)^{-1})}{T_{j-1}(\mu-\mathcal{L}_0)}\label{e:UB}\\
\mathcal{H}_{2j+2} &\gtrsim \frac{1}{C^{2j+2}}(-\mathcal{L}_0) T_{j}(\mu-\mathcal{L}_0)\label{e:LB}
\end{align}
(see~\cite[Theorem 3.3]{CET} for the precise statement), where the inequalities above are to be intended 
in the sense of operators, $C>1$ is a constant, uniformly bounded from below for $\lambda$ small, 
arising from the approximations made in each step of the iteration 
and the function $T_j$ is defined as a  
Taylor expansion truncated at level $j$, i.e. 
\begin{equation}\label{e:T}
T_j(x)=\sum_{\ell=0}^j\frac{(\tfrac12 \log\log(1+x^{-1}))^\ell}{\ell!}\,.
\end{equation}
Plugging~\eqref{e:LB} and~\eqref{e:UB} into~\eqref{e:b}, choosing $\mu$ sufficiently small 
and $j$ sufficiently large depending on $\mu$ and $C$ ($j\approx C^{-2}\log\log (1/\mu)$),~\eqref{maineqL} follows with $\delta \approx 1/C^2$.\\
To give a taste of the computations involved, let us show how to derive \eqref{e:UB} for $j=1$. Testing $\mathcal H_3$ against a $n$-particle state $\phi\in \Gamma L^2_n$ and using the explicit expression for $\mathcal A_\pm$, we get
\begin{equation}
\begin{aligned}
  \langle \phi,\mathcal H_3\phi\rangle_n\sim &\lambda^2\int {\mathrm d} k_{1:n} |k_{1:n}|^2|\hat \phi(k_{1:n})|^2\\&\times\int {\mathrm d}p \frac{(K_{p,k_1-p})^2}{\mu+\tfrac{|p|^2+|k_1-p|^2+|k_{2:n}|^2}2}
\end{aligned}  
\end{equation}
with $k_{1:n}=(k_1,\dots,k_n)$ and $|k_{1:n}|^2=|k_1|^2+\dots+|k_n|^2$. Using $(K_{p,k_1-p})^2\le1$ and performing the integral on $p$, one obtains an upper bound of the form
\begin{equation}
 \lambda^2 \int {\mathrm d} k_{1:n} |k_{1:n}|^2|\hat \phi(k_{1:n})|^2\log\left(1+(\mu+ \tfrac{|k_{1:n}|^2}2)^{-1}\right)
\end{equation}
which implies \eqref{e:UB} for $j=1$.  Let us remark that already this
first bound implies a divergence of the bulk diffusivity, of order at
least $\log\log t$ for $t$ large. Indeed, plugging it into~\eqref{e:b}
with $j=3$,  one can show that
\begin{equation}
b_3(\mu) \gtrsim \int {\rm d}p \frac{(K_{p,-p})^2}{\mu+\tfrac{|p|^2}{2}(1+\log(1+(\mu+\tfrac{|p|^2}{2})^{-1}))}\,,
\end{equation}
where $|p|\leq 1$.
Carefully evaluating the integral yields $b_3(\mu)\gtrsim \log \log(1/ \mu)$, from which the claim follows at once.
\medskip

\paragraph*{Conclusions.}
We have studied the AKPZ equation, an anisotropic variant  of the $2d$
KPZ equation \eqref{eq:KPZ} with $\det Q<0$, which has a Gaussian,
logarithmically rough stationary state.  The common folklore belief is
that it has the same scaling behaviour as the (linear) EW
equation. While indeed our results confirm that both have the same
(vanishing) roughness and growth exponents, we prove that
non-linearity produces non-trivial logarithmic corrections to the
diffusive scaling and to the bulk diffusion coefficient. In fact, \emph{we
propose these corrections as a distinguishing feature between the EW and the AKPZ
universality classes for $2d$ stochastic growth}. It would be extremely
interesting to find (numerical and/or analytical) evidence of analogous
logarithmic super-diffusivity for discrete growth models as those in
\cite{borodin2009anisotropic,MR3936124,BT}, that are conjectured to have
the same qualitative features as the AKPZ equation, or for the
equation with non-linearity given by
$[(\partial_{x_1}H)^2-a (\partial_{x_2}H)^2], a>0$. In fact, for
$a\ne 1$ the stationary state is \emph{not Gaussian}
\cite{da2003nonlinear} but the RG analysis of \cite{W91} suggests
that the behaviour should be the same as for $a=1$, in particular the
stationary state should be asymptotically Gaussian on large scales, as
indicated by the simulations in \cite{Healy}. Even though log corrections to
diffusivity may look too tiny to be observed, we emphasise that
the predicted $(\log t)^{2/3}$-effect in $2d$ driven diffusive models
has been very recently numerically measured \cite{Krug}.

This work was partially supported by EPSRC grant EP/S012524/1, by National Council for Scientific and Technological Development - CNPq via a 
Universal grant 409259/2018-7, and a Bolsa de Produtividade 303520/2019-1 and by ANR-15-CE40-0020-03 Grant LSD.
\bibliography{bibtex}

\end{document}